\documentclass[reprint,pra]{revtex4-2}
\usepackage{graphicx}
\usepackage{bm}

\def\opex{Opt. Express}
\def\josaa{J.\ Opt.\ Soc.\ Am. A}
\def\ol{Opt.\ Lett.}
\def\prl{Phys.\ Rev.\ Lett.}

\begin{document}

\title{Complex modes in optical fibers and silicon waveguides}
\author{Nan Zhang}
\affiliation{Department of Mathematics, City University of Hong Kong,
  Hong Kong, China}
\author{Ya Yan Lu}
\affiliation{Department of Mathematics, City University of Hong Kong,   Hong Kong, China}
\date{\today}

\begin{abstract}
  At a fixed frequency, a general wave field in an open optical waveguide
  is  a  sum of finite number of guided modes and an integral
  of radiation modes. Most studies on open lossless dielectric
  waveguides are concerned with regular guided modes which are confined around
  the waveguide core, have a real propagation constant, and propagate along
  the waveguide axis without attenuation. However, there may be guided
  modes with a complex propagation constant even though the waveguide 
  consists of lossless isotropic dielectric materials, and they are
  the so-called complex modes. The complex 
  modes are proper modes confined around the
  waveguide core, and are different from the leaky modes which have
  divergent wave fields in the transverse plane. For 
  waveguides with discontinuities and defects, the complex modes can
  be excited and must be included in eigenmode expansions. However, 
  existing studies on complex modes in open lossless dielectric waveguides are
  very limited. In this Letter,  we consider circular 
  fibers and silicon waveguides,  study the formation mechanism of complex
  modes, and calculate the dispersion relations for
  several complex modes in each waveguide.  For circular fibers, we
  also determine the minimum refractive-index ratio for the existence
  of complex modes. Our study fills a gap in optical waveguide theory
  and provides a basis for realizing potential applications of complex
  modes. 
\end{abstract}

\maketitle

\section{Introduction}

For open lossless dielectric waveguides, a regular guided mode at a
given real frequency is a proper eigenmode with a field confined
around the waveguide core~\cite{snyder,vassallo}. It has a real
propagation constant, carries a finite power, and propagates along the
waveguide axis without attenuation. 
If the waveguide is lossy, any guided mode must have a complex
propagation constant due to absorption loss, and it propagates forward 
with an exponentially decaying amplitude.  A leaky mode
\cite{snyder,vassallo,aop}, on the other 
hand, is an improper mode with a divergent field in the transverse
plane. It radiates out power in lateral directions, has a complex
propagation constant due to radiation loss, and also decays
exponentially as it propagates forward. The so-called complex modes
also have a complex propagation constant, but not because of 
absorption and radiation losses~\cite{mrozo97}.
Instead, the appearance of complex propagation 
constant is related to the non-self-adjoint linear operators 
in full-vector waveguide eigenvalue formulations for a given real  
frequency.
Complex modes were first discovered in closed waveguides
supporting backward waves~\cite{tai60,cla60,cla64}.
The existence of complex modes in open lossless dielectric
waveguides was first demonstrated by Jablo\'nski for a high
index-contrast optical fiber~\cite{jablo94}. It has been shown that complex modes also exist in silicon
waveguides and they are related to the numerical instability of the full vectorial
paraxial beam propagation  method~\cite{xie11}. In addition, complex modes also
exist in periodic waveguides~\cite{amgad} and in waveguides with anisotropic media or
metamaterials~\cite{jiang09,nari19}. 
Guided modes with complex propagation constants also appear 
in ${\cal PT}$-symmetric waveguides where the real and imaginary parts 
of the complex refractive index 
profile are even and odd (in a transverse variable) 
respectively~\cite{ruter10,gold18}, but 
they are different from the complex modes in lossless waveguides. 
Unlike the leaky modes, a complex mode in an open waveguide is a proper
eigenmode with a field confined around the core. A unique property
of complex modes is that they do not transport power along the waveguide axis.
For waveguides with discontinuities or a local defects,
complex modes can be excited and must be included in
eigenmode expansions. 

The existing studies on complex modes in open lossless dielectric
waveguides are very limited. The example of Jablo\'nski is for a
circular fiber with a very high index ratio ($\rho = n_1/n_2 = 7.81$, where
$n_1$ and $n_2$ are the refractive indices of the core and the
cladding, respectively)~\cite{jablo94}. For silicon waveguides, complex modes are
only calculated for a single frequency~\cite{xie11}. In this Letter,
we carry out a more detailed study on complex modes in circular fibers
and silicon waveguides. We show that complex modes form bands, and  each
band is given by a propagation constant 
being a complex-valued function of the real frequency.
For circular fibers,
we also determine the minimum values for index ratio $\rho = n_1/n_2$ beyond
which complex modes of different azimuthal order appear.

\section{Theory}

\begin{figure}[h]
  \centering
    \includegraphics[scale=0.5]{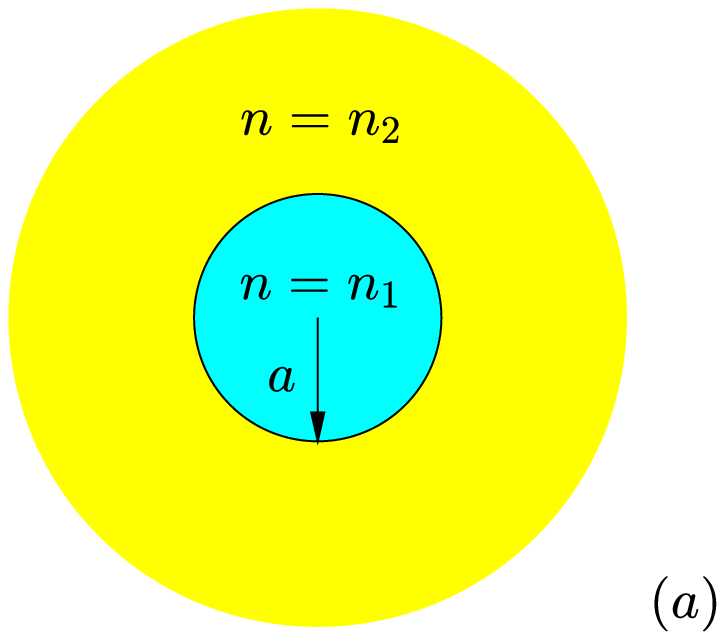}\quad
  \includegraphics[scale=0.5]{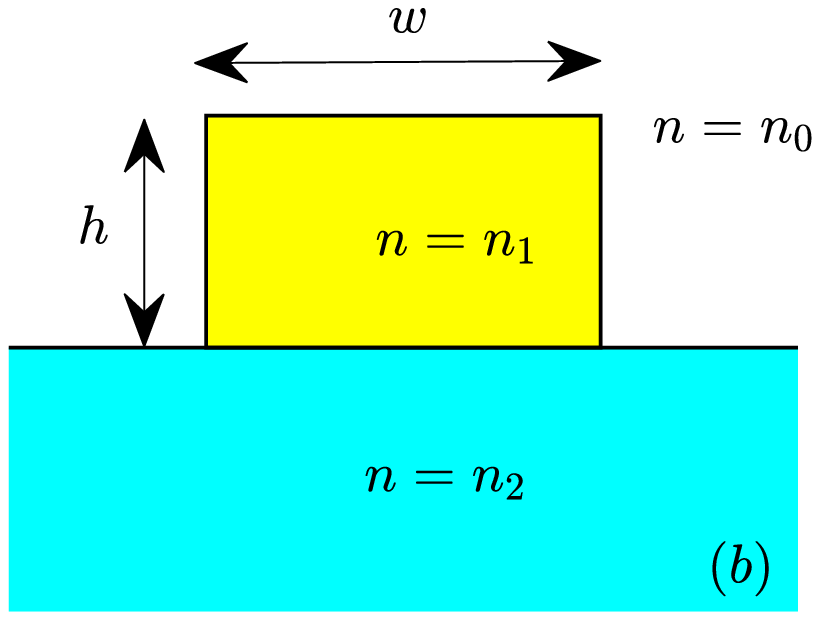}
  \caption{(a) A circular optical fiber of radius $a$ and refractive index $n_1$
    with a cladding of refractive index $n_2$. (b) A silicon waveguide
    with a rectangular cross section (width $w$, height $h$) on a
  silica substrate. The refractive indices of the core, the substrate
    and the cladding are $n_1$, $n_2$ and $n_0$, respectively.}
  \label{twoguides}
\end{figure}

We consider a circular step-index optical fiber with a homogeneous cladding and a
silicon waveguide on a silica substrate,  and depict  their cross sections in
Fig.~\ref{twoguides}. 
Let $z$ be the variable along the waveguide axis, $x$ and $y$ be the
transverse variables,  we consider eigenmodes that depend on $z$
and time $t$ as $e^{ i (\beta z - 
  \omega t)}$, where $\omega$ is the angular frequency and $\beta$
is the propagation constant. We are concerned with the eigenvalue
formulation where $\omega$ is real and given, and $\beta$ (or $\beta^2$) is the
eigenvalue~\cite{snyder,vassallo}. 
An eigenmode is designated as a proper one, if the
field decays exponentially to zero as
$r = \sqrt{x^2 + y^2} \to \infty$. For a lossless waveguide, the
regular guided modes are proper eigenmodes with a real 
$\beta$. They form bands with starting
points on the light line $\omega = (c/n_2) |\beta|$ and are given by
their dispersion relations $\omega = \omega_{j}(\beta)$, where
$c$ is the speed of light in vacuum and $j$
is an integer index. 
For the two waveguides shown in Fig.~\ref{twoguides}, it appears
that all dispersion curves of the regular guided
modes have a positive slope for a positive $\beta$, i.e., $d
\omega_{j}/d\beta > 0$ for $\beta > 0$.

A leaky mode in an open waveguide must satisfy an outgoing radiation
condition and decay to zero as it propagates forward (to $z =  +
\infty$). Therefore, the complex propagation constant $\beta$ of a
leaky mode (depending on $z$ as $e^{ i \beta z}$) should satisfy
$\mbox{Re}(\beta) > 0$ and $\mbox{Im}(\beta) > 0$. 
For both waveguides shown in Fig.~\ref{twoguides}, in the
unbound medium with refractive index $n_2$,  all electromagnetic field
components of the eigenmode satisfy the same Helmholtz equation
\begin{equation}
  \label{helom}
  \partial_x^2 u + \partial_y^2 u + (k_0^2 n_2^2 - \beta^2) u = 0, 
\end{equation}
where $k_0 = \omega/c$ is the freespace wavenumber. 
For the complex $\beta$ of a leaky mode, $k_0^2 n_2^2 - \beta^2$ is in
the lower half of the complex plane, $\kappa_2 = \sqrt{ k_0^2 n_2^2 -
  \beta^2}$, by the standard complex square root function with a branch cut along the
negative real axis, is in the fourth quadrant, then the outgoing
radiation condition implies that
\begin{equation}
  \label{farfield}
  u \sim \frac{C_0(\theta)}{\sqrt{r}} e^{ i \kappa_2 r}, \quad r \to \infty, 
\end{equation}
where $\theta$ is the polar angle of $(x,y)$ and $C_0$ is a function
depending on $\theta$. Since $\mbox{Im}(\kappa_2) < 0$, $u$ 
diverges as $r \to \infty$, thus a leaky mode is an improper
mode with a divergent field in the transverse plane. 

The complex modes found in closed waveguides many decades ago emerge from local
maxima on the dispersion curves of regular guided
modes~\cite{tai60,cla64}. Some complex 
modes in periodic waveguides (with a periodicity along the waveguide
axis) also emerge from such points where the group
velocity is zero~\cite{amgad}. This implies
that that the associated dispersion curve (of regular guided modes)
must contain a region with a 
negative slope corresponding to backward
waves~\cite{cla60}. However,  backward waves do not seem to
exist in open lossless $z$-invariant waveguides consisting of ordinary 
isotropic dielectric materials. Therefore, a different mechanism is
needed to realize complex modes in open lossless waveguides. It turns
out that a leaky mode simply turns to a complex mode when the real part
of the propagation constant $\beta$ passes zero. More specifically, if
the originally positive $\mbox{Re}(\beta)$ of a leaky mode becomes
negative as $\omega$ is varied, and $\mbox{Im}(\beta)$ remains 
positive, then $k_0^2 n_2^2 - \beta^2$ moves to the upper half plane,
$\kappa_2$ moves to the first quadrant, and the field component $u$, 
satisfying condition (\ref{farfield}), decays to zero as $r \to
\infty$. Therefore,    a leaky mode simply becomes a complex mode when
$\mbox{Re}(\beta)$ changes sign.  Since the propagation constant
$\beta$ of the leaky mode is in the first quadrant, that of the
complex mode is in the second quadrant.
An open optical 
waveguide typically has an infinite sequence of leaky modes, and each 
leaky mode satisfies a dispersion relation with $\beta$ being a
complex-valued function of $\omega$~\cite{aop}. If on a particular dispersion 
curve, there is an interval of $\omega$ in which $\mbox{Re}(\beta) <
0$, then, the eigenmode on that interval is actually a complex mode. 

From the reciprocity principle, if the waveguide has a mode with
propagation constant $\beta$, it must have another one with
propagation constant $-\beta$. This implies that the propagation
constants of the leaky and complex modes can also lie in the third and
fourth quadrants, 
respectively. For a lossless waveguide with a real refractive index
profile, we can take a complex conjugate for a complex mode (with
propagation constant $\beta = \beta' + i \beta''$)  satisfying
the governing Maxwell's equations, and obtain another complex mode
with propagation constant $\overline{\beta} = \beta' - i
\beta''$. Together with the reciprocity, we conclude that complex
modes appear in a group of four, with propagation constants
$\beta'  \pm  i \beta''$ and  $-\beta'  \pm i \beta''$. If we take
complex conjugate for a leaky mode (with propagation constant $\beta$)
satisfying the Maxwell's equations,  we obtain an eigenmode 
with propagation constant $\overline{\beta}$, but it is not a leaky
mode, since the outgoing radiation condition is reversed. The
obtained mode may be called absorbing mode, since it gains power from lateral
incoming waves and propagates forward with an exponentially increasing
amplitude. In summary, a pair of leaky modes (with propagation constants in  first
and third quadrants, respectively) give rise to two complex modes with
propagation constants in the second and fourth quadrants, and a pair of
absorbing modes give rise to the other two complex modes (with
propagation constants in  first and third quadrants).

\section{Circular fiber}

For the step-index circular fiber shown in Fig.~\ref{twoguides}(a), the eigenmodes
are classified into groups depending on the azimuthal order $\nu$.
For a leaky or complex mode with a propagation constant $\beta$ in the
upper half plane,  and if the angular dependence is $\sin (\nu 
\theta)$ or $\cos (\nu \theta)$,
then the field components in the cladding are proportional to
$H_\nu^{(1)}(\kappa_2 r) \sin( \nu \theta)$ or
$H_\nu^{(1)}(\kappa_2 r) \cos ( \nu \theta)$,
where $H_\nu^{(1)}$ is $\nu$th order Hankel function of first kind.
The above choice ensures that the asymptotic condition
(\ref{farfield}) is satisfied.

For $\nu = 0$, $\beta$ satisfies the
dispersion equation  
\begin{equation}
  \label{case0}
	\frac{\kappa_2H^{(1)}_0(\kappa_2a)}{H^{(1)}_1(\kappa_2a)}=
        \eta  \frac{\kappa_1J_0(\kappa_1a)}{J_1(\kappa_1a)}, 
      \end{equation}
where $\kappa_1= (k_0^2 n_1^2 - \beta^2)^{1/2}$,  $J_0$ and $J_1$ are Bessel functions of first kind, 
$\eta=1$  and $\eta = \rho^2$ for the TE and TM polarizations,
respectively. It can be proved that Eq.~(\ref{case0}) does not have any
solution with $\mbox{Re}(\beta)=0$. If this is not true, i.e., there is a
$\beta$ such that $\mbox{Re}(\beta)=0$, then $\beta^2 \le 0$, the
right hand side of Eq.~(\ref{case0}) is real, but the imaginary part
of the left hand side is nonzero, because 
\begin{eqnarray*}
  \mbox{Im}\left[
          \frac{H^{(1)}_0(\kappa_2a)}{H^{(1)}_1(\kappa_2a)}\right]
  &=& \frac{Y_0(\kappa_2a)J_1(\kappa_2a)-J_0(\kappa_2a)Y_1(\kappa_2a)}{J_1^2(\kappa_2a)+Y_1^2(\kappa_2a)}\\
  &=& \frac{ 2/(\pi \kappa_2 a)} {J_1^2(\kappa_2a)+Y_1^2(\kappa_2a)}
      \ne 0, 
\end{eqnarray*}
where $Y_0$ and $Y_1$ are Bessel functions of second kind. In fact, for $\nu=0$, the eigenmodes are 
separately in the TE and TM polarizations, the related eigenvalue
problems are self-adjoint, and complex modes do not exist. 

\begin{figure}[h]
  \centering
  \includegraphics[scale=0.29]{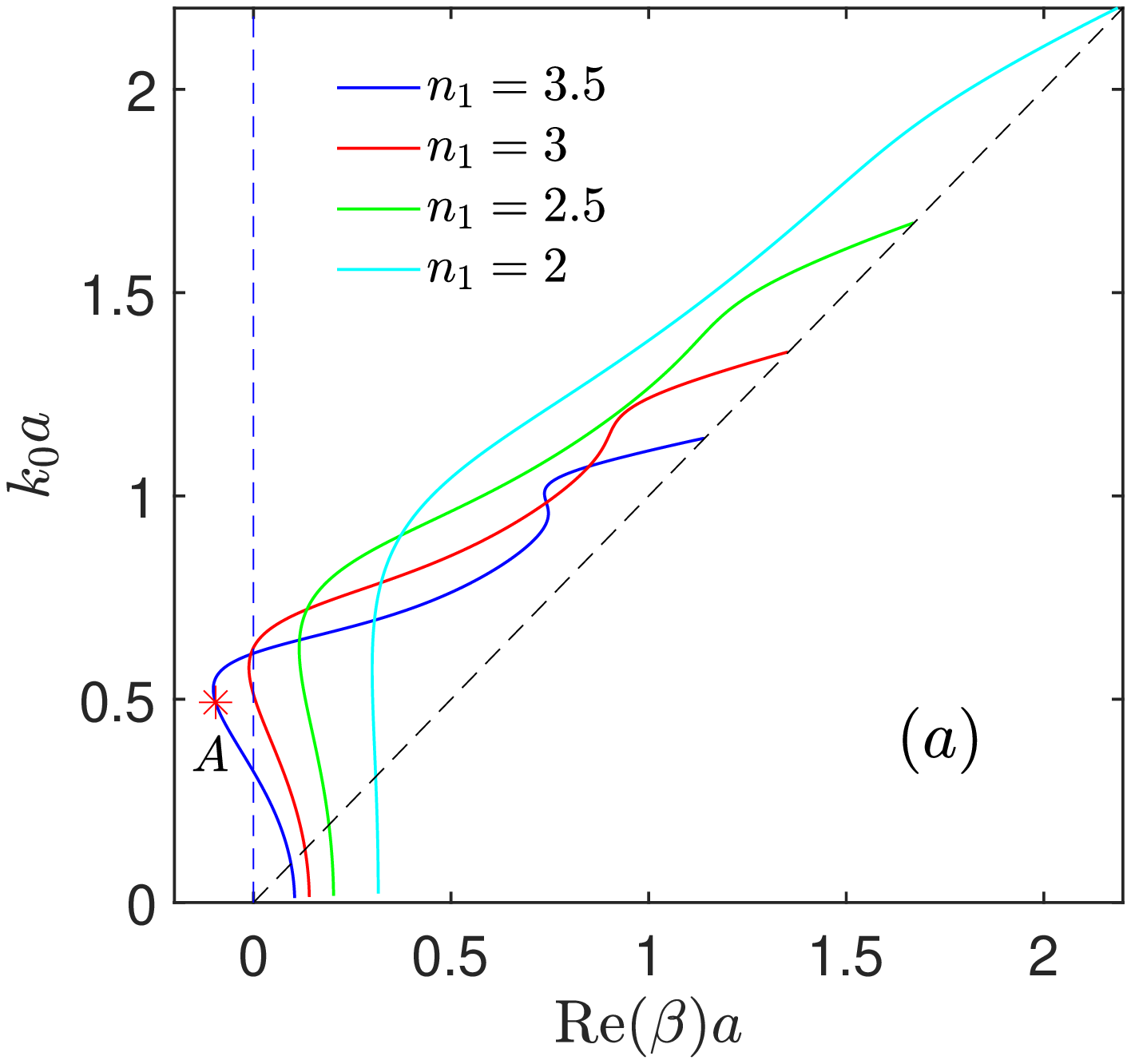} 
  \includegraphics[scale=0.29]{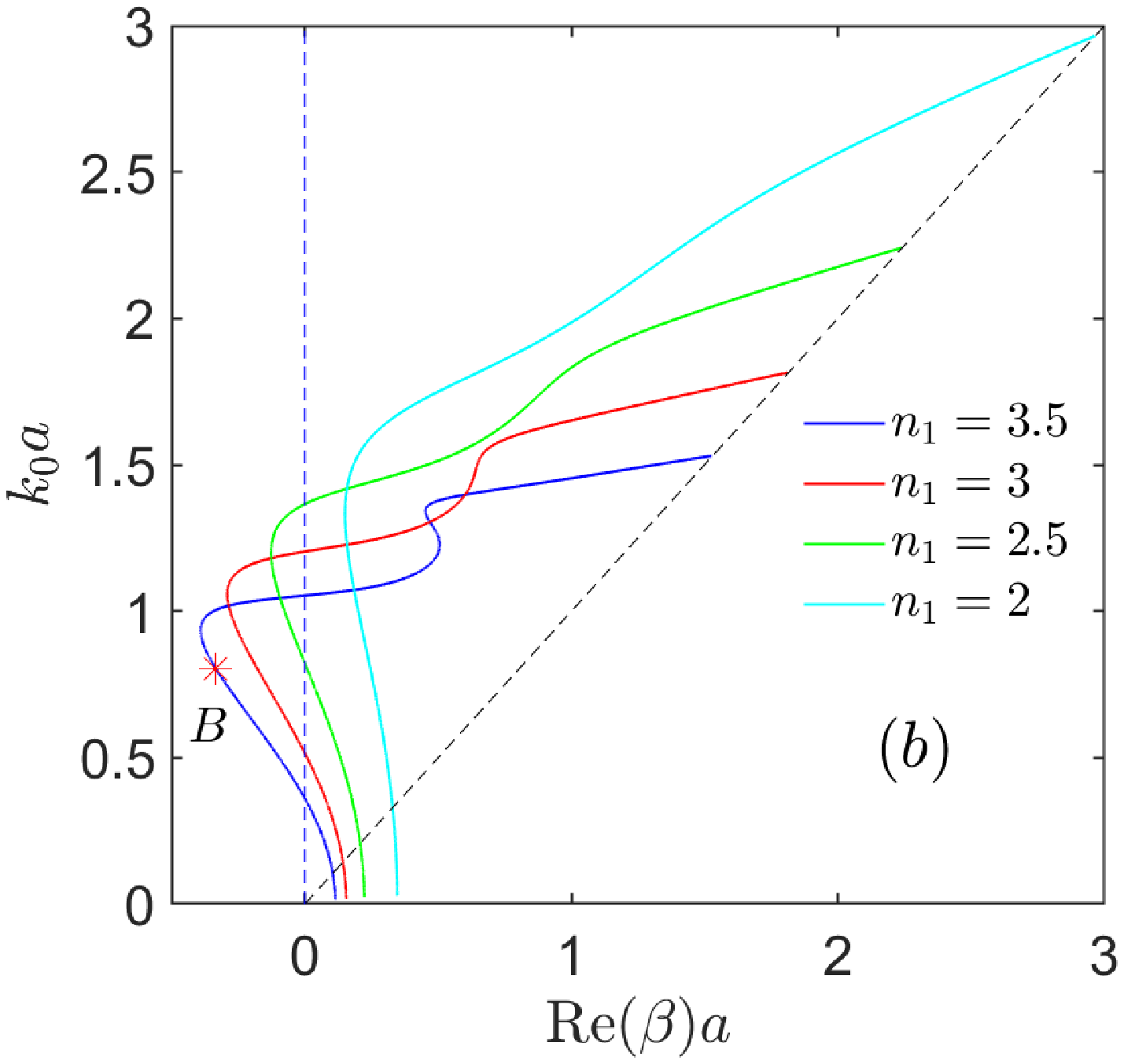}
  \includegraphics[scale=0.3]{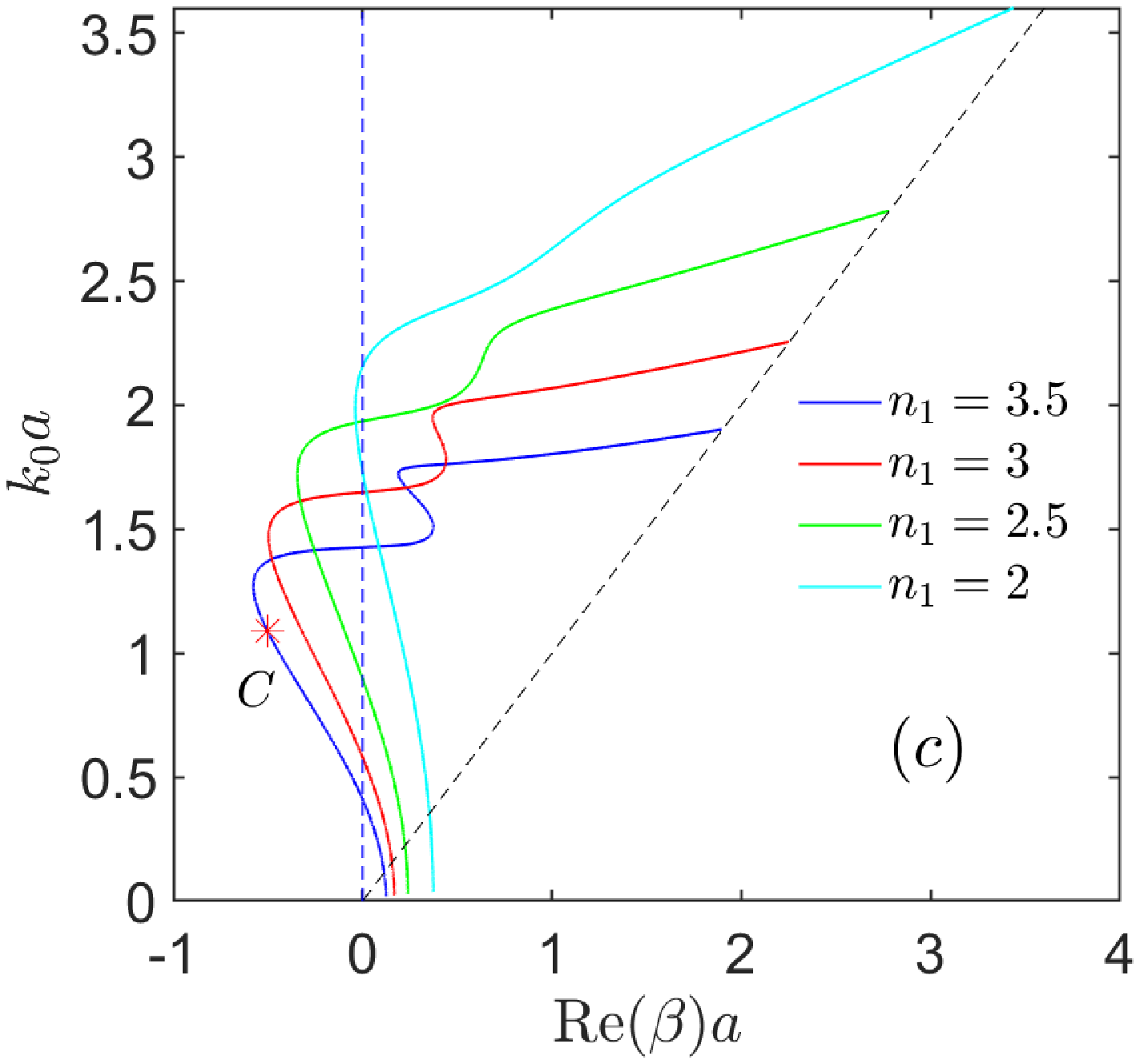}  
  \includegraphics[scale=0.3]{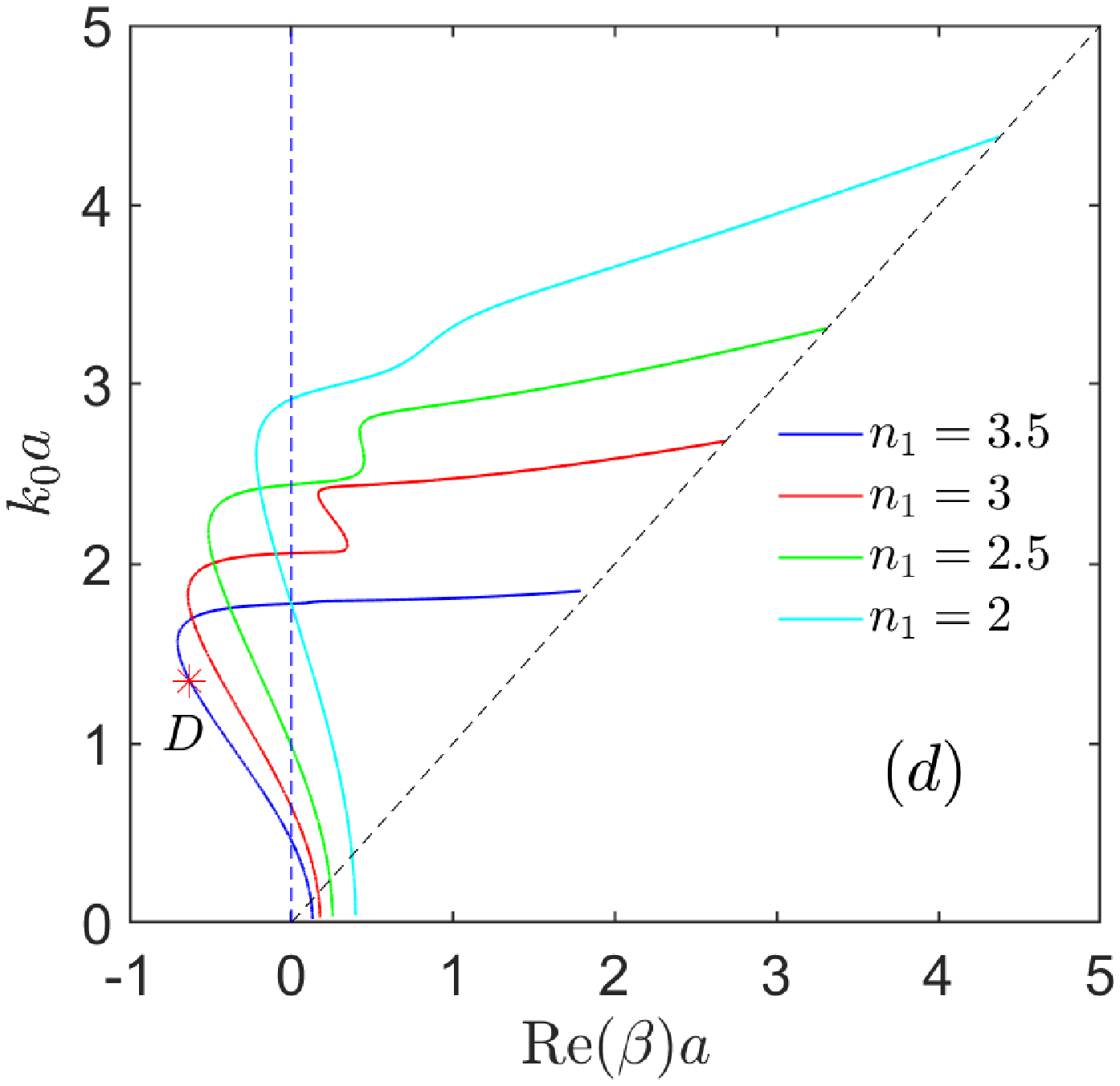}
  \caption{Dispersion curves [$k_0$ vs. $\mbox{Re}(\beta)$] of complex
    and leaky modes in a circular fiber of radius $a$ surrounded by air ($n_2 =
    1$). Panels (a), (b), (c) and (d) correspond to azimuthal order
    $\nu=1$, 2,  3, and 4, respectively. The refractive index of 
  the core is assumed to be $n_1 = 2$, 2.5, 3 or 3.5.}
  \label{curvemu}
\end{figure}

\begin{figure}[h]
  \centering 
  \includegraphics[scale=0.3]{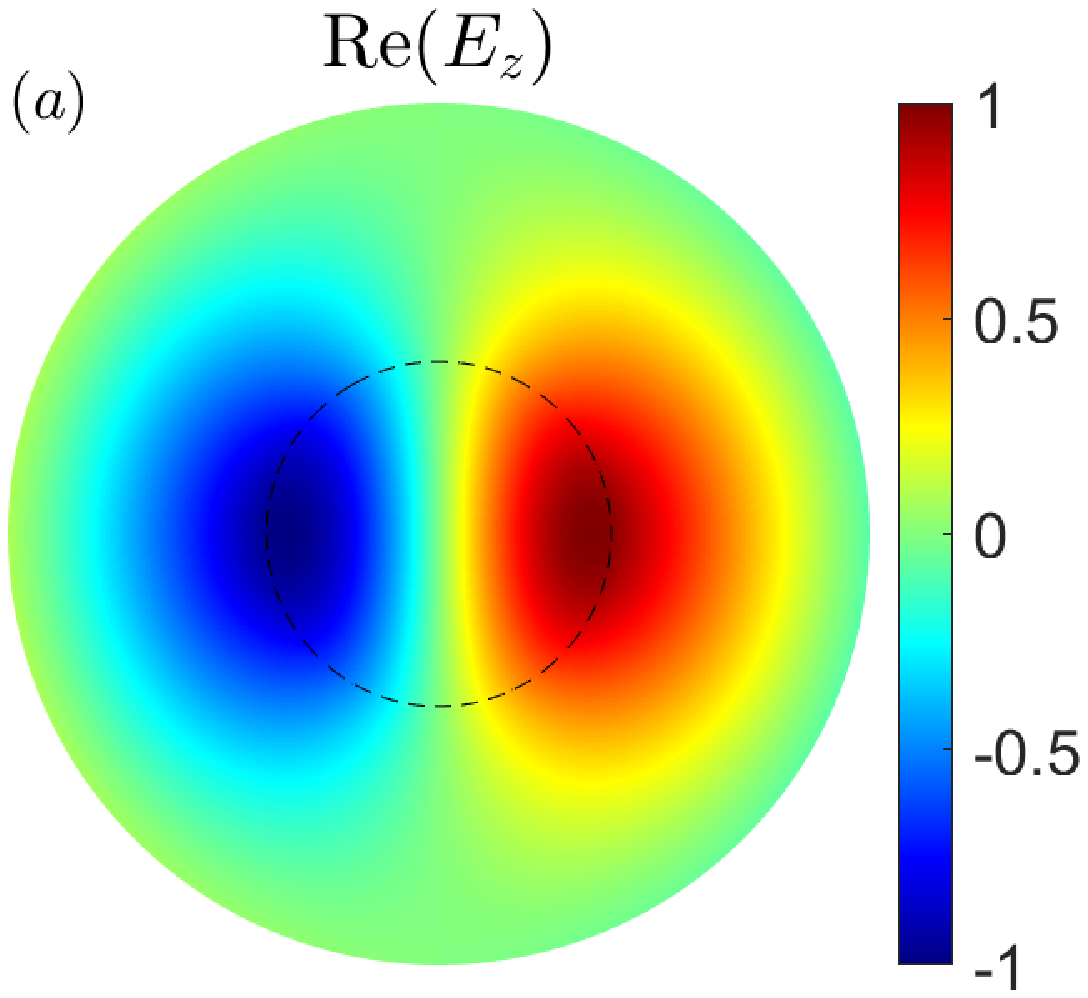} \quad
  \includegraphics[scale=0.3]{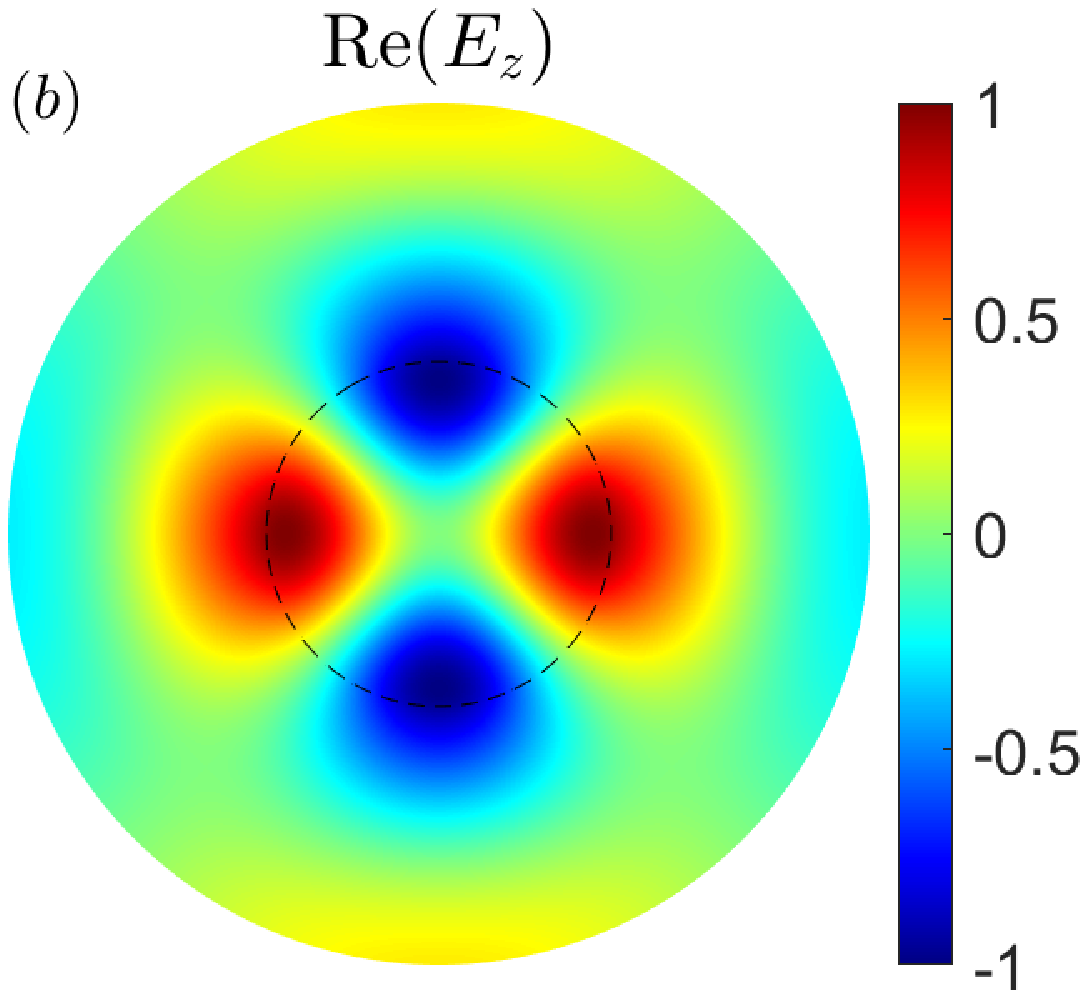}
  \includegraphics[scale=0.3]{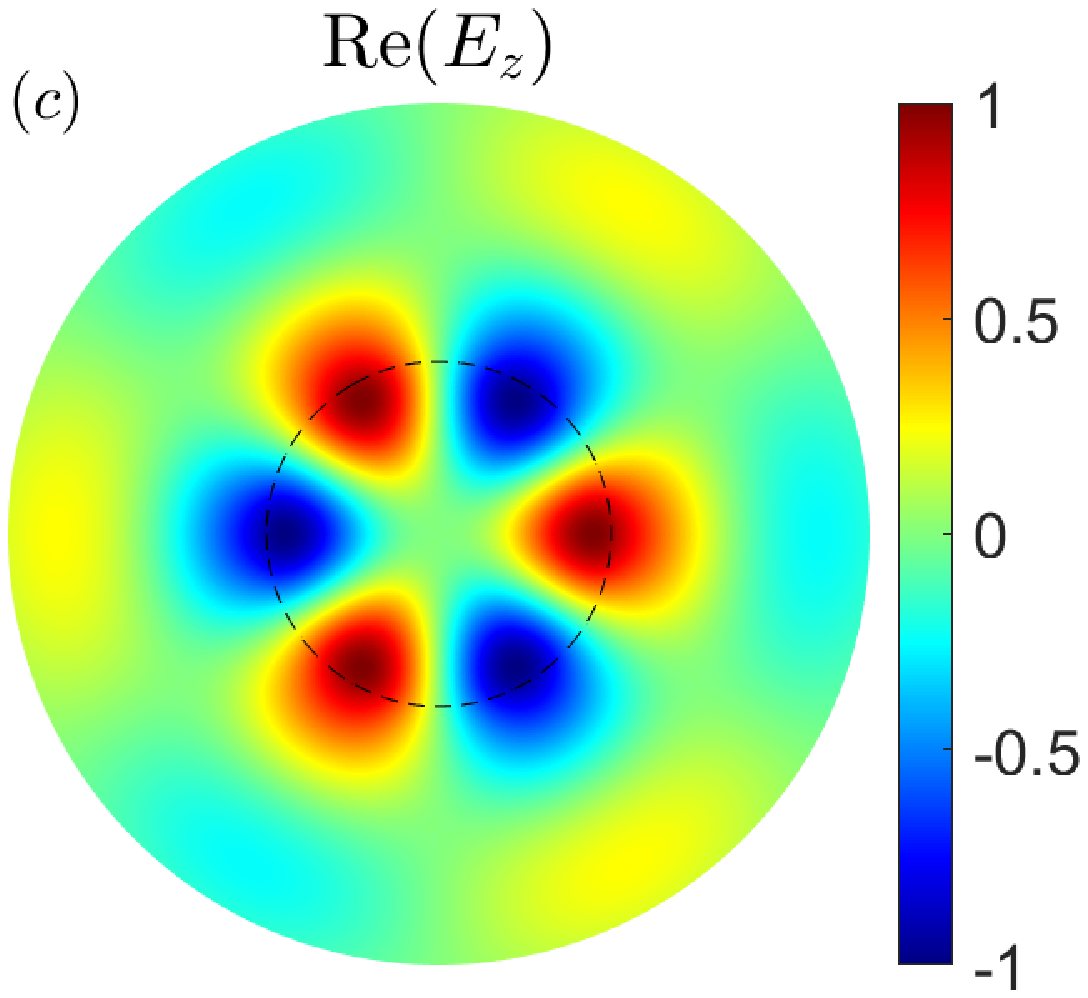} \quad 
  \includegraphics[scale=0.3]{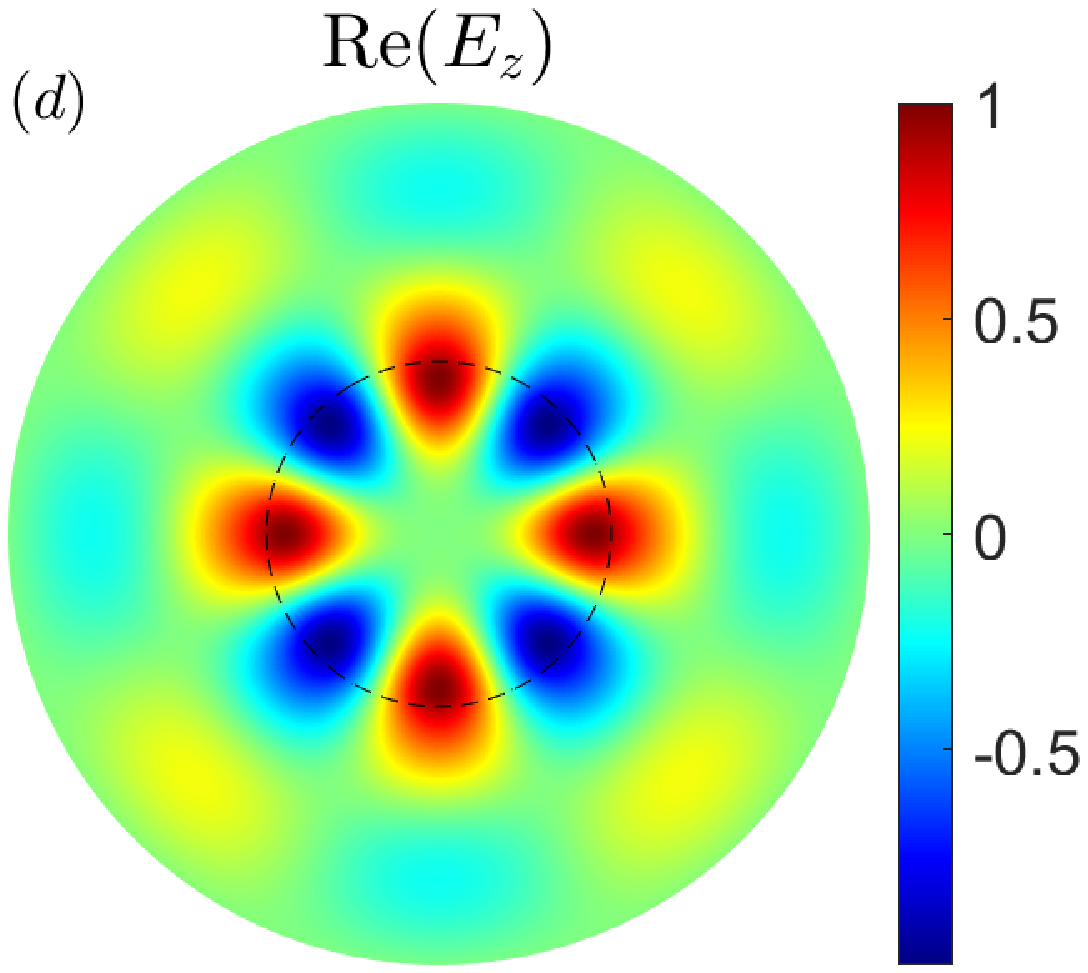}
  \caption{Field patterns (real part of $E_z$) of complex modes in an 
    optical fiber with $n_1=3.5$ and $n_2=1$. (a) $\nu=1$,
    $k_0a=0.4925$, $\beta a = -0.0958 + 1.1512i$.  
    (b) $\nu=2$,   $k_0a=0.8055$, $\beta a = -0.3338 + 1.9785i$. 
    (c) $\nu=3$,   $k_0a=1.0895$, $\beta a = -0.5021 + 2.8037i$. 
    (d) $\nu=4$,   $k_0a=1.3502$, $\beta a = -0.6290 + 3.6455i$.}
  \label{ezmu}
\end{figure}

For $\nu \ne 0$,  $\beta$ can be solved from the following
dispersion equation
\begin{equation}
  \label{dispeq}
  (\rho^2 J^{-} - H^{-})(J^+ - H^+) = 
  (H^+ - \rho^2 J^{+} )(J^- - H^-),
\end{equation}
where 
\[
  J^{\pm} = \frac{ J_{\nu\pm 1} (a\kappa_1) }
  { a\kappa_1  J_\nu (a\kappa_1) },
  \quad 
  H^{\pm} = \frac{H^{(1)}_{\nu\pm 1} ( a \kappa_2) }
  { a\kappa_2  H^{(1)}_\nu (a\kappa_2) }.
\]
Jablo\'{n}ski found a complex mode for a very large index ratio $\rho =
n_1/n_2 = 7.81$~\cite{jablo94}. However, complex modes do exist in
optical fibers with  lower index contrast. In Fig.~\ref{curvemu}(a), (b), (c) and (d), 
we show the dispersion curves [freespace wavenumber $k_0$
vs. $\mbox{Re}(\beta)$] of some leaky and complex modes for $\nu=1$, 2, 3 and 4, respectively. The results
are obtained for a fixed $n_2 =1 $ and $n_1 = 2$, 2.5, 3 and 3.5. The
leaky and complex modes are shown together in each curve for positive
and negative $\mbox{Re}(\beta)$, respectively. It can be seen that
complex modes exist when $\rho = 3$ and 3.5 for $\nu=1$,
when $\rho = 2.5$, 3 and 3.5 for $\nu=2$, and when
$\rho = 2$, 2.5, 3 and 3.5 for $\nu=3$ and 4. In Fig.~\ref{ezmu},
we show the $z$ component of the electric field for four complex modes
marked as $A$, $B$, $C$ and $D$ in
Fig.~\ref{curvemu}. The azimuthal dependence is assumed to be $\cos (\nu \theta)$.

\begin{figure}[b!]
  \centering
  \includegraphics[scale=0.46]{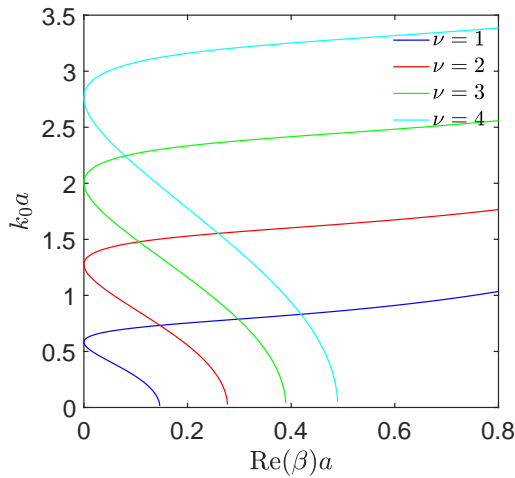}
  \caption{Dispersion curves of leaky modes in a circular fiber with
    index ratio $\rho = \rho_\nu$ for azimuthal 
    order $\nu=1$, 2, 3, 4.}
  \label{critical}
\end{figure}

The above numerical results suggest that complex modes can exist in the optical
fiber for each $\nu \ne 0$ whenever the index ratio $\rho$ is larger
than a critical value $\rho_\nu$.  For $\rho = \rho_\nu$,
$\mbox{Re}(\beta)$, as a function of $k_0$, has a minimum which is
exactly zero. We can calculate the critical index ratio $\rho_\nu$, $k_0$ and
$\mbox{Im}(\beta)$ at the minimum from a system consisting of Eq.~(\ref{dispeq}) and
\begin{equation}
  \label{redbeta}
  \mbox{Re} \left( \frac{ d \beta}{ dk_0} \right) = 0.
\end{equation}
The results are listed in Table~\ref{table1}. 
\begin{table}[h]
  \centering
  \caption{Critical values of index ratio $\rho = n_1/n_2$ for leaky mode
    branches reaching a zero minimum of $\mbox{Re}(\beta)$.}
  \begin{tabular}{c|c|c|c} \hline
    $\nu$ & $\rho_\nu$ & $n_2 k_0a$ &  $\mbox{Im}(\beta) a$ \\ \hline
    1 & 2.9474 &   0.5827 &   1.1513 \\ \hline
    2 & 2.2396 &   1.2753 &  1.8171  \\ \hline
    3 & 1.9574 &  2.0060 &  2.4092 \\ \hline
    4 & 1.7985  & 2.7620 &  2.9626 \\ \hline
  \end{tabular}
  \label{table1}
\end{table}
In Fig.~\ref{critical},
we show four dispersion curves of the leaky modes for $\rho = \rho_\nu$ and $\nu =
1$, 2, 3 and 4, respectively. 

\section{Silicon waveguide}

\begin{figure}[h]
      \centering 
      \includegraphics[scale=0.39]{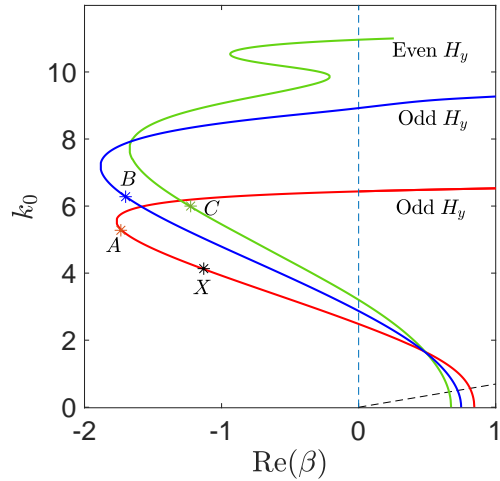}
      \caption{Three dispersion curves [$k_0$ vs. $\mbox{Re}(\beta)$]
        of complex and leaky modes in the silicon waveguide shown in
        Fig.~\ref{twoguides}(b). The waveguide parameters are
        $w=500$\,nm, $h=250$\,nm, $n_1=3.479$, $n_2=1.445$,  and
        $n_0=1$. The unit for both horizontal and vertical axes is $1/\mu$m.}
      \label{siliband}
    \end{figure}

\begin{figure}[h!]
   \centering 
   \includegraphics[scale=0.285]{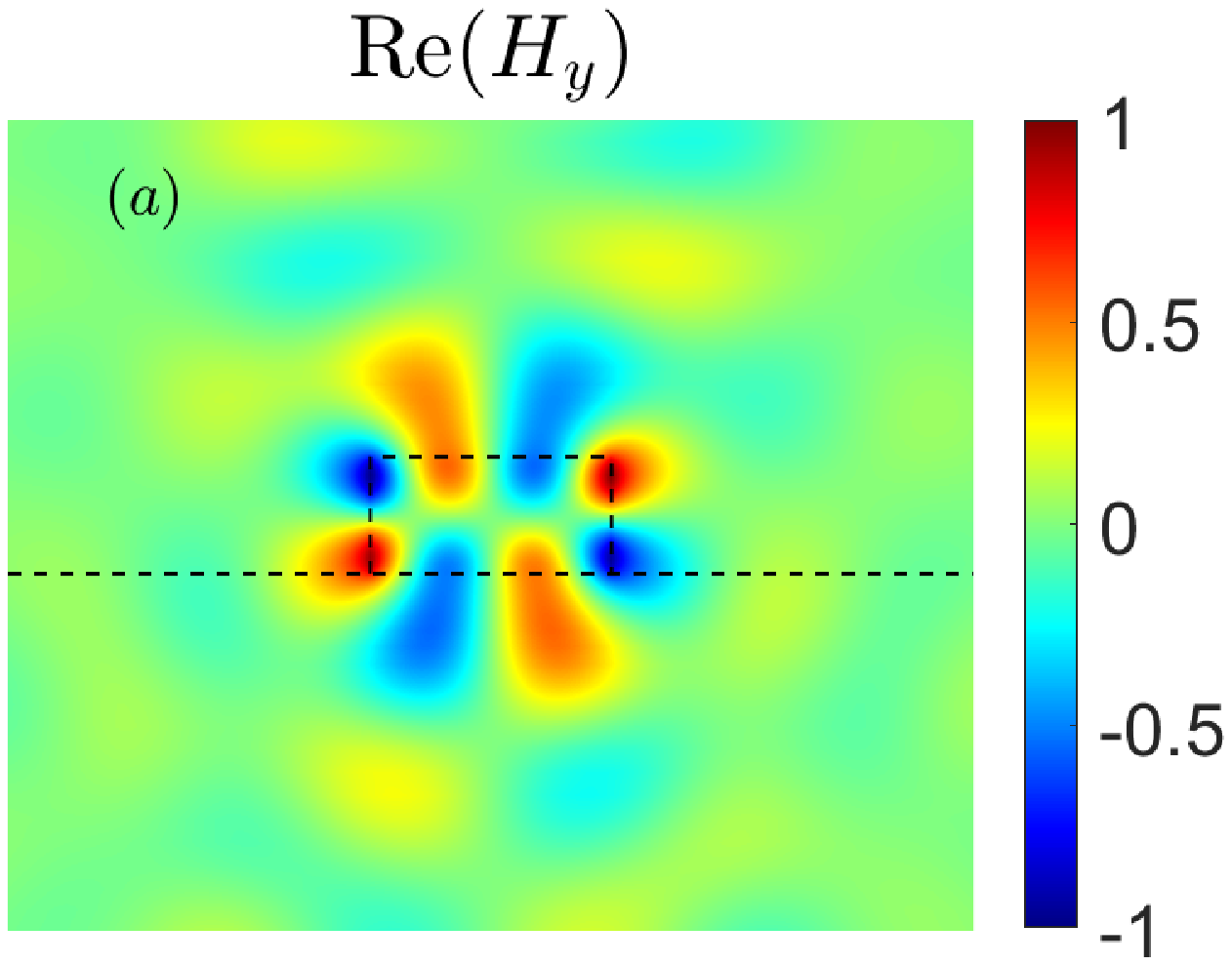}
   \includegraphics[scale=0.285]{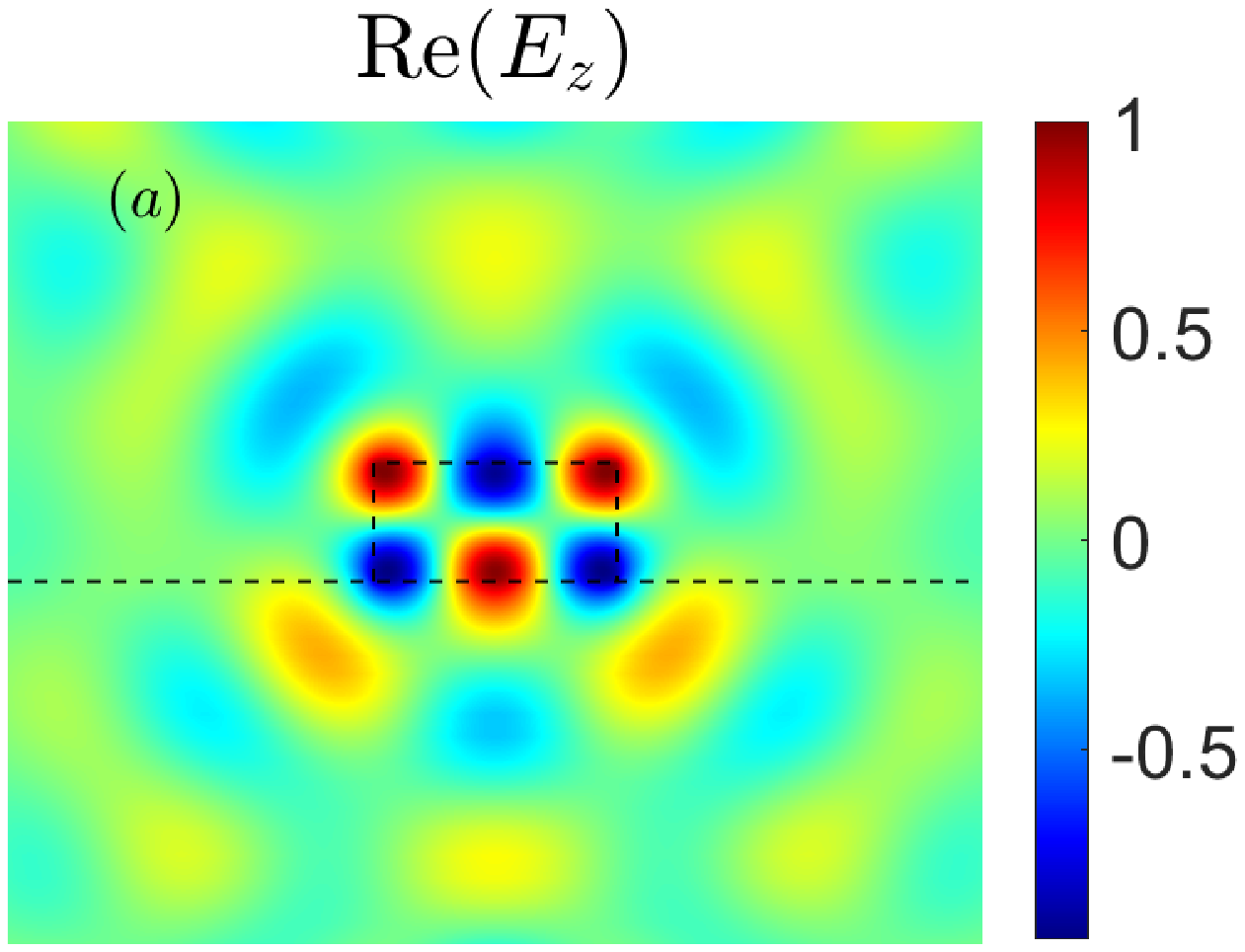}
   \includegraphics[scale=0.285]{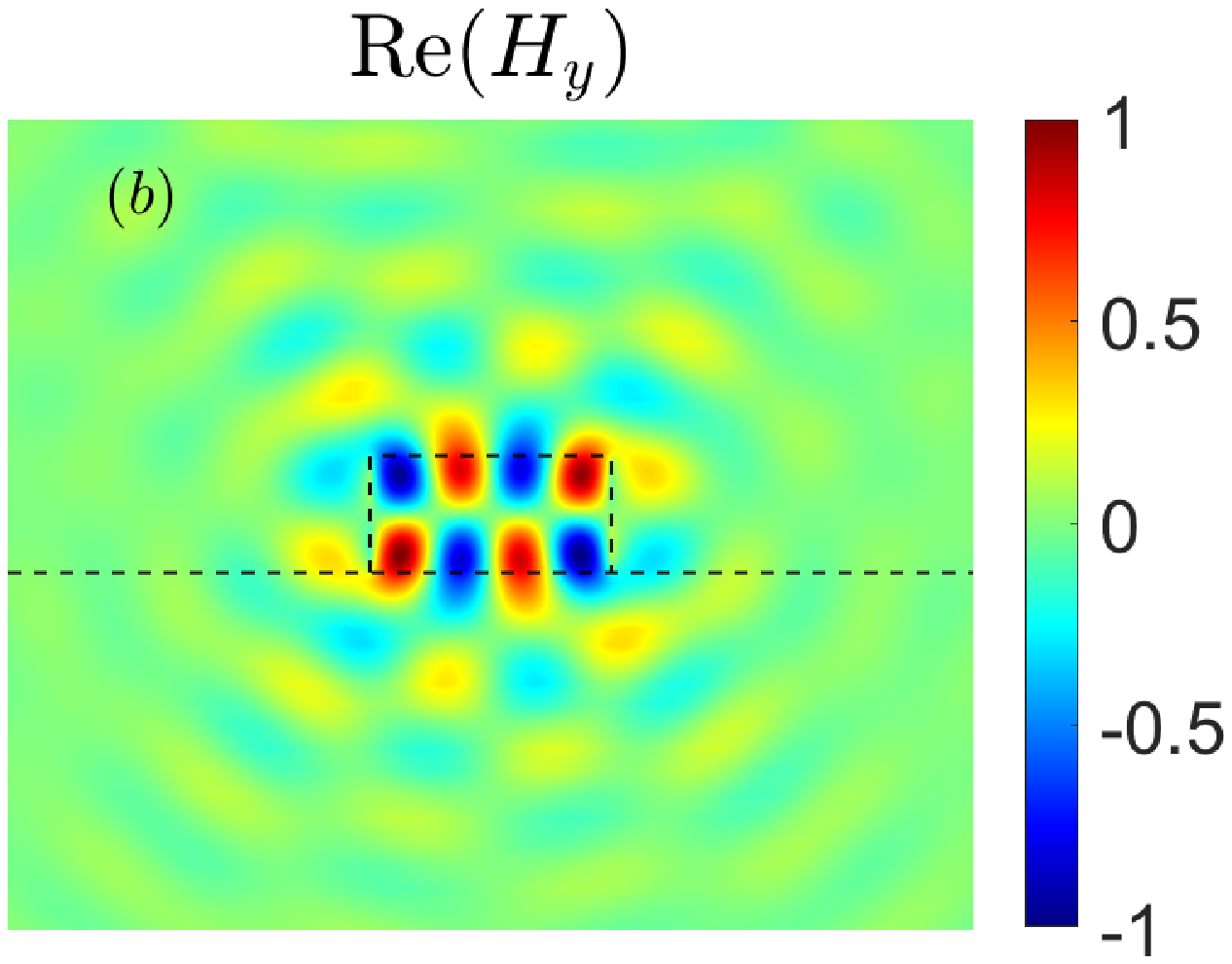}
   \includegraphics[scale=0.285]{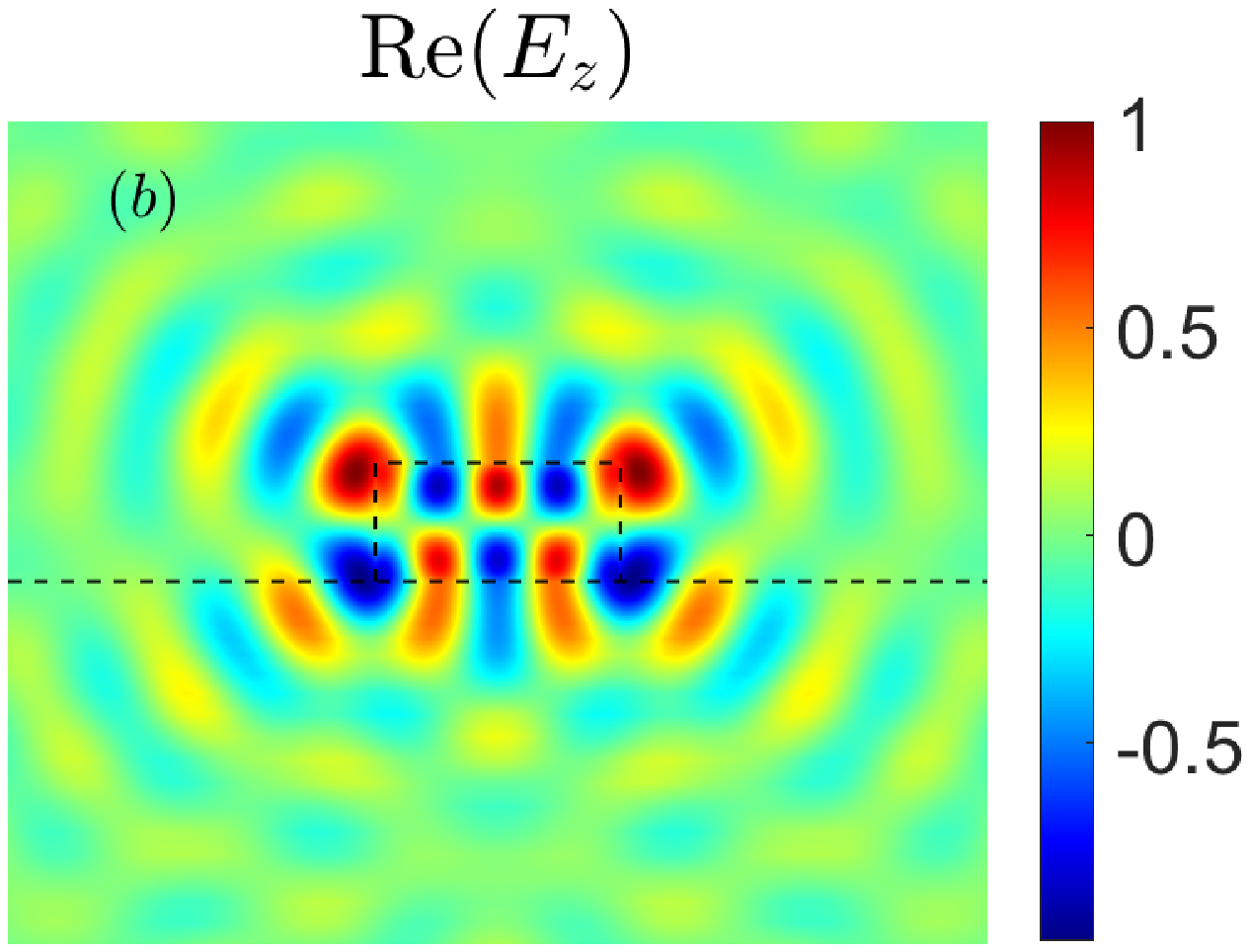}
   \includegraphics[scale=0.28]{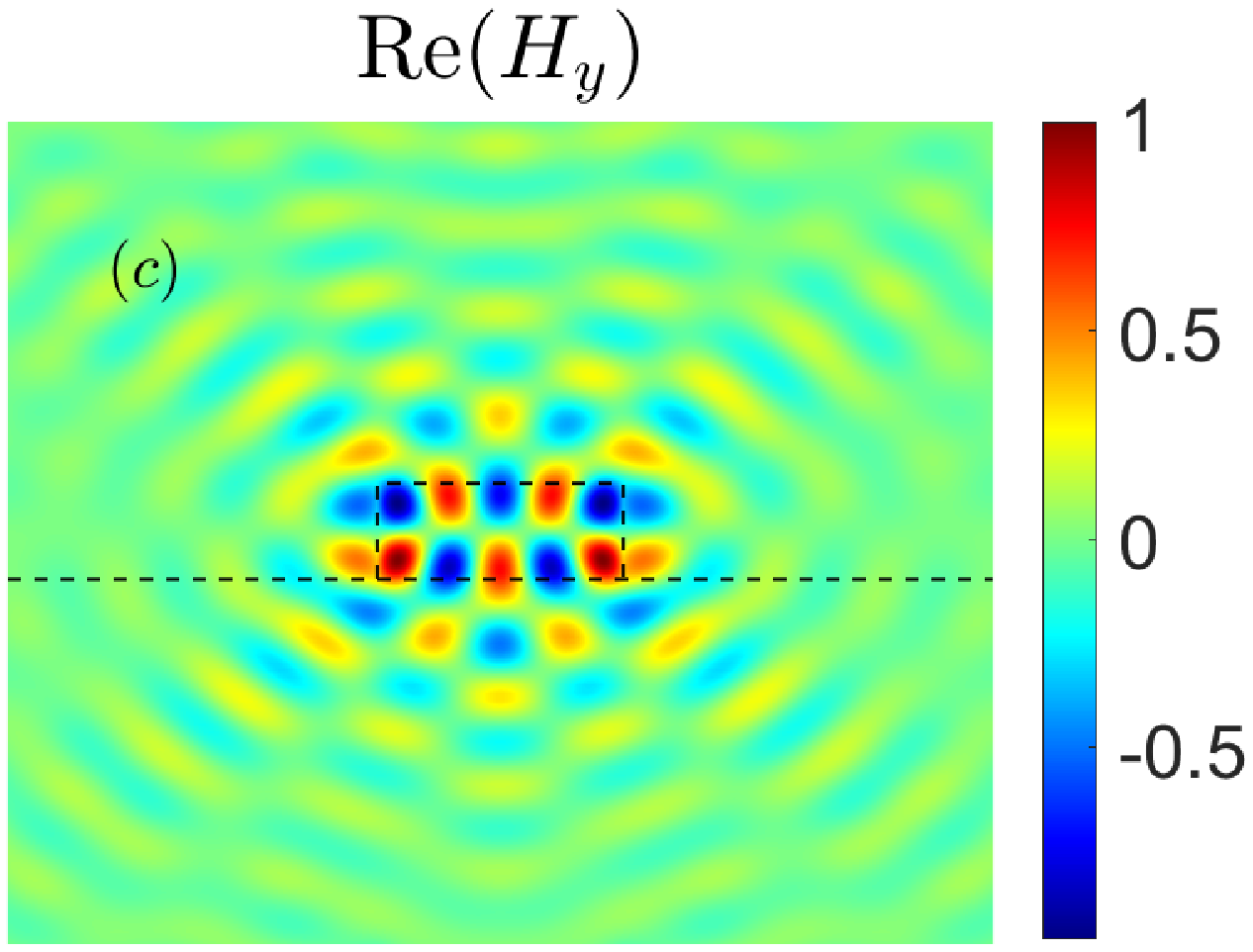}
   \includegraphics[scale=0.28]{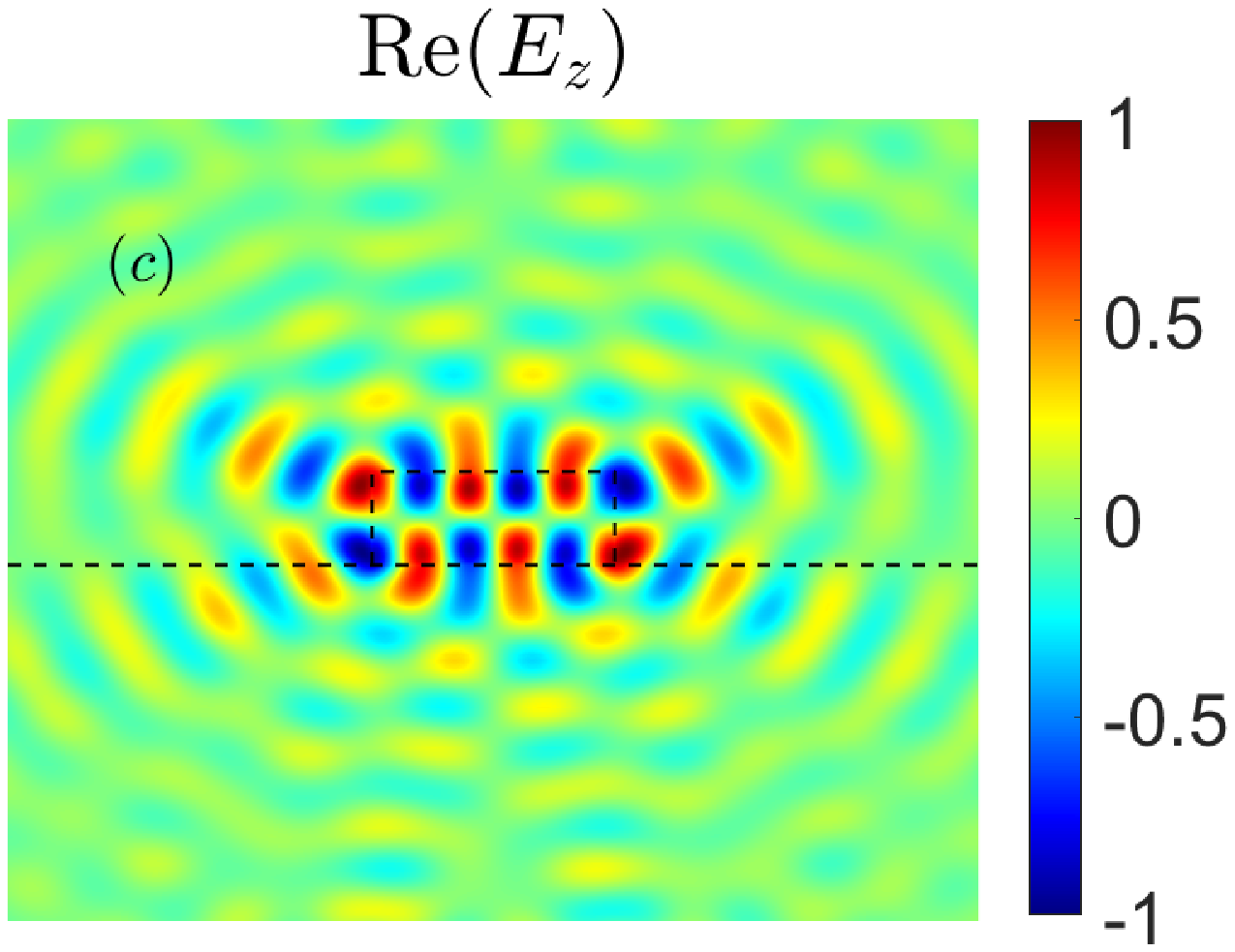}
   \caption{Wave field patterns of selected complex modes in a silicon
     waveguide, marked as 
     $A$,  $B$ and $C$ in Fig.~\ref{siliband}. 
     (a) $k_0=5.2779$, $\beta  = -1.7348 + 11.1339i$;
     (b) $k_0=6.2793$, $\beta  = -1.7219 + 22.0920i$;
     (c) $k_0=6.0000$, $\beta  = -1.2256 + 29.2028i$. The unit for
     $k_0$ and $\beta$  is  $1/\mu\mbox{m}$.}
   \label{silifield}
 \end{figure}

In \cite{xie11}, the silicon waveguide is assumed to have a rectangular
cross section with width $w = 500$\,nm, height $h=250$\,nm, and
refractive index $n_1 = 3.479$,  and is placed on a substrate with
refractive index $n_2 = 1.445$ and covered by air ($n_0=1$). A complex mode was
calculated for freespace wavelength $\lambda_0 = 1.52\,\mu$m.  Similar
to the case of the optical fiber, 
a complex mode is formed when a leaky mode reaches
$\mbox{Re}(\beta) = 0$ at some frequency. In Fig.~\ref{siliband}
 we show three dispersion curves of leaky and complex modes, where 
 $\mbox{Re}(\beta)$ passes through zero. For each dispersion curve,
 the complex modes exist in an interval of $k_0$ in which
 $\mbox{Re}(\beta) < 0$. The example obtained by Xie {\it et al.}~\cite{xie11}
 corresponds to point $X$ in Fig.~\ref{siliband}. In
 Fig.~\ref{silifield}, we show field patterns for three complex modes
 marked as $A$, $B$, $C$ in Fig.~\ref{siliband}. 

 \section{Conclusion}
 
 In summary, a complex mode in an open lossless dielectric waveguide
is a proper eigenmode with a field confined around the waveguide core, has
a complex propagation constant due to the non-self-adjoint nature of
the waveguide eigenvalue
problem formulated at a given real frequency, and it does not carry a
power along the waveguide 
axis. We have calculated the dispersion curves for some complex modes
in circular optical fibers and silicon waveguides, and found the
minimum index ratio for the existence of complex modes of various
azimuthal order in circular fibers. Unlike the  complex modes in
closed waveguides that 
emerge from local maxima on dispersion curves of regular guided
modes, the complex modes studied in this Letter are obtained from
leaky modes when the real part of the complex propagation constant
changes sign. While the study of optical waveguides has a long
history, complex modes in open lossless dielectric waveguides are
seldom mentioned in existing literature. Our study fills a gap in
optical waveguide theory. It is worthwhile to further analyze the
properties of the complex modes and explore their potential applications.

\vspace{0.4cm}
\noindent {\bf Funding.} The Research Grants Council of Hong Kong 
Special Administrative Region, China (Grant No. CityU 11304619).

\vspace{0.4cm}
\noindent {\bf Disclosures.} The authors declare no conflicts of
interest.

\end{document}